\documentclass[aps,amsmath,showpacs,twocolumn]{revtex4-1}

\usepackage{graphics,epsfig}

\begin{document}


\title{The inelastic vs. total nucleon-nucleon cross section at large $N_c$}

\author{Thomas D. Cohen}
\email{cohen@physics.umd.edu}

\author{Vojt\v{e}ch Krej\v{c}i\v{r}\'{i}k}
\email{vkrejcir@umd.edu}

\affiliation{Maryland Center for Fundamental Physics, Department of Physics, \\
 University of Maryland, College Park, MD 20742-4111}

\begin{abstract}
In this paper, the implication of the extreme large $N_c$ limit for nucleon-nucleon cross section is examined.
Starting from the $N_c$ scaling of the S-matrix elements, 
a relation between the total inelastic and total cross sections is derived for the regime in which the momenta are much larger than the QCD scale.    
A conceptual complication arise from the fact that there is a tower of  baryons, such as the $\Delta$, with mass splittings from the nucleon which go to zero at large $N_c$.    
Since these baryons are stable at large $N_c$, the meaning of elasticity must be modified.
Processes which only transform nucleons to these baryons are considered elastic; 
processes in which at least one additional meson is produced are considered inelastic.  
It is shown that in the extreme large $N_c$ limit,  the total inelastic cross section is exactly one eighth of the total cross section. 
In contrast, in the physical world of $N_c=3$ the total cross section for high-energy nucleon-nucleon scattering is dominated by inelastic processes.


\end{abstract}


\maketitle

The approach to quantum chromodynamics (QCD) based on an expansion around the limit of infinite number of colors
has attracted wide interest since its introduction in the 70s \cite{tHooftNC, WittenNC}. 
Its attractiveness 
stems from the fact that many aspects
of the QCD simplify in the limiting case of $N_c \rightarrow \infty$.   
Moreover, to the extent that physical world of $N_c=3$ is close to the infinitely colorful world of $N_c \rightarrow \infty$, 
one can obtain useful phenomenological predictions following this approach.  
Many qualitative and semi-quantitative predictions are possible since certain quantities are analytically
calculable in the leading order of the $1/N_c$ expansion.  
This is remarkable since at present there is no known way to solve QCD even at leading order in the ordinary expansion in powers of the coupling constant.  
For example, it can be shown that an emergent spin-flavor symmetry 
\cite{GervaisSakita84-1,GervaisSakita84-2,DashenManohar93-1, DashenManohar93-2,DashenJenkinsManohar94,DashenJenkinsManohar95}
arises for certain baryon observables at large $N_c$, and this symmetry allows for predictions of {\it ratios} of various physical quantities with corrections which are known to be of relative order $1/N_c$ or in favorable cases $1/N_c^2$ \cite{DashenJenkinsManohar94, DashenJenkinsManohar95, CohenKrejcirik18}.   
The predictions of these tend to describe the physical world remarkably well \cite{JenkinsLebed}.

Recently, there has been  interest in the implications of the large $N_c$ limit
on nucleon-nucleon scattering.  
The spin-flavor dependence of the inclusive differential cross section---that is the cross section for the nucleons to emerge at a fixed angle with any number of mesons produced---was deduced in
Ref. \cite{Banerjee***} for the kinematic regime in which the incident momenta are large compared to $\Lambda_{\rm QCD}$ (or more precisely for the regime of Witten kinematics \cite{WittenNC} in which the incident velocity of the nucleons is held fixed when $N_c$ is taken large).  
Since the mass of the nucleon is proportional to $N_c$, this automatically yields kinematics with large momenta---even 
large compared to the $\Lambda_{\rm QCD}$ scale. 
The analysis of  Ref.~\cite{Banerjee***} follows from the emergent symmetry discussed above.  
It was argued in Ref.~\cite{CohenGelman12} that the total cross section should follow the same spin-flavor dependence as the inclusive differential cross section.

The development of an analytic formula for the total nucleon-nucleon 
cross section in the extreme large $N_c$ limit (in which $\log(N_c) \gg 1$) in the Witten kinematics,
which was done in Ref.~\cite{CohenPRL2012}, 
represents an important theoretical breakthrough.  
The result is
\begin{equation}
\sigma_{\rm tot} = \frac{2 \pi}{m_{\pi}^2} \log^2(N_c) \,. \label{resulttot}
\end{equation}  
A  general analysis of the assumptions underlying the derivation of Eq.~(\ref{resulttot}) was presented in Ref.~\cite{CohenKrejcirik};
the key issue in that work is the $N_c$ scaling of the real and imaginary parts of nucleon-nucleon S-matrix elements.  
It was argued that the proper quantity to focus on is the logarithm of the S-matrix.  
A number of arguments strongly suggests 
that both the real and imaginary parts of the logarithm of the S-matrix elements are proportional to $N_c$.  
The present work is a follow up to these papers; a new interesting result is derived determining a fraction of cross section corresponding to inelastic processes---again restricted to Witten kinematics and to 
the extreme large $N_c$ limit.  In particular, we find the following relation to hold:
\begin{equation}
\frac{\sigma_{\rm in}}{\sigma_{\rm tot}} = \frac{1}{8} \,. \label{resultratio}
\end{equation}

There are two  important caveats to this result.  
The first is that at large $N_c$, there is a tower of baryons with spin equal to isospin which is degenerate with the nucleon; more specifically, with the mass splitting proportional to $1/N_c$.  
Thus, reactions in which nucleons only transform to other members of this degenerate multiplet of states and {\it no}
other particles (mesons) are emitted 
are considered elastic at large $N_c$. This definition of elasticity is natural
since the transition from nucleon to other baryon requires no extra energy since they are degenerate.
The second caveat is that, like Eq.~(\ref{resulttot}), this formula only holds in the extreme large $N_c$ limit in which  $\log(N_c) \gg 1$. This is not a computational problem, but rather a problem if one thinks about phenomenological relevance of obtained relations.  
Note that the world of $N_c=3$ is far from this limit.  Thus, the result---while 
of considerable theoretical interest---should not be expected to hold even approximately at $N_c=3$. Indeed it does not.

First, let us briefly summarize the basic properties  of nucleon physics
in the  large $N_c$ limit,
and their consequences for our analysis.
We also use this opportunity to set the notation used in this paper.
According to the standard $N_c$ scaling rules \cite{WittenNC}, the mass of the nucleon is
proportional to $N_c$. The Witten kinematic regime, which has a smooth large $N_c$ limit, is
the one for fixed velocity; thus the nucleon momentum is also of order $N_c$.
The self-consistency of a Schr\"odinger type  equation in this regime requires also the strength
of the interaction to be of order $N_c$ \cite{KaplanSavage, KaplanManohar} 
while the range of the interaction is to be kept finite.
In summary:
\begin{equation}
M= N_c \tilde{M} \,\,,\,\,\, k= N_c \tilde{k} \,\,,\,\,\, V(r) = N_c \tilde{V}(r) \,\,.
\label{basicscaling}
\end{equation}
Characters with tilde indicate the variables with the $N_c$ scaling removed; plain characters stand
for the full quantities---this notation is used throughout the paper.  
It directly follows from Eq.~(\ref{basicscaling}) that the angular momentum $l$ is also proportional to $N_c$ if one considers the
semi-classical limit, in which the impact parameter $b$ is held fixed. Recall that the semi-classical limit is appropriate at large $N_c$ and Witten kinematics \cite{CohenPRL2012, CohenKrejcirik}. 

The key theoretical quantity determining the scattering process is the elastic S-matrix, where the meaning of 
elastic is defined as above.  
The incident and final states depend on the spin and flavor configurations of the baryons;  
let us label one such configuration by $A$.  The 2-body states also depend on the angular momentum $l$.  
Thus a 2-body elastic S-matrix element is specified by $S_{A,l;A'l'}^{\rm elastic}$.  We note that in the collision 
of nucleons the total angular momentum $j$ rather than orbital one $l$ is conserved; so the S-matirx is not diagonal in $l$.
This technicality, however, does not alter the leading $N_c$ behavior discussed in this work.

The limit in which we are interested is $N_c \rightarrow \infty$. This limit considered in the kinematic regime specified by 
scaling (\ref{basicscaling}) 
corresponds to the semi-classical approximation to quantum mechanics.  
Of course the actual problem of interest, 
the nucleon-nucleon scattering at large momentum, is inherently a field-theoretical problem---it is possible to create other 
particles (mainly mesons).  However, this can be captured in a quantum mechanical language by considering a potential with nonzero imaginary part which is matched to the results of field theory.
On the one hand, the imaginary part leads to a violation of unitarity and a loss of flux;
on the other hand, it is precisely what one expects during inelastic processes---the flux is lost from an elastic channel into other channels
(for example a state with nucleons and pions). Since we are not interested in the specifics of inelastic channels, the use of
the quantum mechanics with a complex potential is sufficient. 
The appropriateness of this approach was discussed more extensively in Ref. \cite{CohenKrejcirik}.

The starting point of our discussion is the observation that the logarithm of the S-matrix for the elastic processes is proportional
to $N_c$ in the large $N_c$ limit \cite{CohenKrejcirik} for $k$, $l$ and $l'$ of order $N_c$.  A precise way to state this is that eigenvalues of the elastic S-matrix can be written as  real and imaginary phase shifts.
For our purposes, it is useful to write the $j^{\rm th}$ eigenvalue of the S-matrix, $s_j(\tilde{k})$, in terms of the respective phase shifts:
\begin{eqnarray}
s_j (\tilde{k})&=& \exp \left( - 2 \delta_j^{\rm Im}(\tilde{k}) + 2 i \, \delta_j^{\rm Re}  (\tilde{k}) \right)  \nonumber \\
    &=& \exp \left( - 2 N_c \, \tilde{\delta}_j^{\rm Im} (\tilde{k})+ 2 i N_c \, \tilde{\delta}_j^{\rm Re}(\tilde{k}) \right) \,,
\label{Smatrix}
\end{eqnarray}
where the indices ${\rm Im}$, and ${\rm Re}$ indicate the imaginary and real part of
the phase shift, respectively.  The imaginary part of the phase shift parameterizes the loss of flux into inelastic channels.

The quantity of interest in this work is the inelastic cross section.  We consider the case in which the initial nucleons are in some configuration that we label $A$.
In a partial wave expansion, the general expression for inelastic scattering is given by
\begin{eqnarray}
\sigma_{\rm in}^A &=& \frac{\pi}{k^2} \sum\limits_{l} (2l+1) \left( 1 - \sum_{l',A'} \left| S_{l,A;l',A'}^{\rm elastic}\right|^2 \right) \nonumber \\
&\approx & \frac{\pi}{k^2} \int {\rm d}l^2 \left( 1 - \sum_{l',A'} \left| S_{l,A;l',A'}^{\rm elastic}\right|^2\right) \,\,,
\label{sigmainelastic}
\end{eqnarray}
where the second form is valid in the regime where many partial waves contribute---as they do at large $N_c$ in Witten kinematics.  
Inserting Eq.~(\ref{Smatrix}) and changing variables to an integration over an impact parameter $b=l/k$ yields
\begin{equation}
\sigma_{\rm in}^A= \pi \, \int \, {\rm d}b^2 \left( 1 - \sum_j  \left | v^j_{A , N_c b \tilde{k}}   \right|^2\  e^ {-4 N_c \, \tilde{\delta}_j^{\rm Im} (\tilde{k}) ) } \right )   \,\,, \label{form1}
\end{equation}
where $v^j_{A , N_c b \tilde{k}} $ is the $A, l=N_c \tilde{l}$ component of the $j^{\rm th}$ normalized eigenvector of the elastic $S$ matrix.  Since they are normalized,
\begin{equation}
 \sum_j  \left | v^j_{A , N_c b \tilde{k}}  \right|^2=1 \,\,.
\end{equation}

In the formal large $N_c$ limit, the factor in parenthesis in Eq.~(\ref{form1}) is equal to one for all values of $b$.  Since the region of integration in $b^2$ extends to infinity, this implies that the inelastic cross-section diverges at large $N_c$. 
In a sense this is not surprising since classical cross section is always infinite unless potential has a strictly finite support;
and the large $N_c$ regime corresponds to semi-classical one.
The interesting question is how the cross section diverges.  
We know that, for any finite value of $N_c$ the cross-sections are finite.  
The reason for this concerns an ordering of limits.  
At sufficiently large impact parameter $b$, the effective absorptive potential goes to zero while at sufficiently large $N_c$ it goes to infinity.  
Thus, for any $b$ at sufficiently large $N_c$, the factor in parenthesis in Eq.~(\ref{form1}) is arbitrarily close to unity.  
However, for any finite $N_c$ at sufficiently large $b$, it is arbitrarily close to zero.   
Thus, at very large but finite $N_c$, one expects the factor to be essentially unity for the region of small $b$ 
and essentially zero when $b\rightarrow \infty$ with a crossover somewhere in between.

For now, let us suppose that the crossover regime is narrow. 
That is, the characteristic width of the crossover region is parametrically small compared to size of the region for which the factor in parenthesis in Eq.~(\ref{form1})   is nearly unity.  
In that case, the  factor can be modeled by a Heaviside step function $\theta(b^2-b_0^2)$ where $b_0$ parameterizes the location
where the crossover takes place---the system looks like a black-disk scattering.   
Thus 
\begin{equation}
\sigma_{\rm in}^A \approx \pi b_0^2 \; . \label{form2}
\end{equation}
It is easy to see that this does in fact happen and that $b_0 \approx \log(N_c)/(2 m_\pi)$ \cite{CohenPRL2012, CohenKrejcirik}.

The key point is that the overall strength of the effective absorptive potential scales as $N_c$ and its longest-ranged 
contribution is of two-pion range \cite{CohenKrejcirik}; it causes an exponential drop off 
which goes as $\exp(-2 m_\pi b)$ times some subexponential function.  
As was shown in Ref.~\cite{CohenKrejcirik} that when the imaginary part of the potential is of order $N_c$ 
then the scaling of 
Eq.~(\ref{Smatrix}) holds.  
On the other hand, when the scale of the imaginary potential is small, the Born approximation for the partial wave 
is valid and the imaginary part of the phase shift is proportional to the imaginary potential itself.  
Clearly at large $N_c$, the transition region is determined by the longest-range part of the interaction since shorter-range 
contributions will exponentiate away 
leaving the interaction strength parametrically of order $N_c$.

One expects the transition region to be centered on a value of $b=b_0$ for which the effective absorptive potential is of order $N_c^0$ since in this region the phase shift are neither parametrically large yielding the factor in parenthesis in Eq.~(\ref{form1}) 
to be effectively unity nor small yielding it close to zero.  
Since the dominant scaling at large $b$ goes exponentially, one expects $b_0$, the center of the transition region, to satisfy
\begin{equation} 
N_c \exp(-2 m_\pi b_0) \sim 1 \; .
\end{equation}
To make this statement more concrete we will solve for $b_0$ using $N_c \exp(-2 m_\pi b_0) = c$, where $c$ is an arbitrary constant of order unity.  
This yields $b_0 = -\log(c)/(2 m_\pi) + \log(N_c)/(2 m_\pi)$.  
Consider the extreme large $N_c$ limit in which $\log(N_c) \gg 1$.  In that case $\log(N_c)/(2 m_\pi)$ dominates 
parametrically over $\log(c)/(2 m_\pi)$ and $b_0$ is well approximated by $\log(N_c)/(2 m_\pi)$.

What is left to be shown is that the width of the crossover region is parametrically small compared to $b_0$ and thus
the approximation using Heaviside step function in  Eq.~(\ref{form1}) is justified.
This is clearly true in the extreme large $N_c$ limit  since  
the width of the crossover region is of order $N_c^0$.  
To see this note that the transition region can be defined as a region in which potential (for a given finite value of $N_c$)
goes from a very large number of 
order unity (yielding large absorption and black disk behavior) to a very small number of order unity (yielding essentially no absorption).  This happens over a few e-folds since the functional form of the potential is exponential; 
one e-fold is characterized by changes of $b$ of order $1/ (2 m_\pi)$, which is independent of $N_c$.  
Thus, the ratio of the width of the crossover region to $b_0$ scales as $1/\log(N_c)$.  In the extreme large $N_c$ limit,
 this is negligible.

Thus, we see that
\begin{equation}
\sigma_{\rm in}^A  =  \frac{ \pi}{4 m_{\pi}^2} \log^2(N_c)
\label{resultinel}
\end{equation}
for any initial spin-flavor state $A$.
Comparing with  Eq.~(\ref{resulttot}) immediately yields Eq.~(\ref{resultratio}).  From the derivation it should be apparent that corrections to Eq.~(\ref{resultratio}) are expected to be of relative order $1/\log(N_c)$.

It is easy to understand why the ratio of the inelastic to the total cross section in Eq.~(\ref{resultratio}) is 1/8.  
Note that the characteristic range of the inelastic scattering is $2 m_\pi$ while for the elastic scattering it 
is $m_\pi$---the longest range elastic interaction is dominated by one-pion exchange whereas inelastic
channels emerge at the level of two-pion exchange.  
Hence the effective $b_0$ for elastic scattering is twice that for the inelastic one. Since the cross section depends on $b_0^2$ 
a factor of 4 emerges in the difference between the total (\ref{resulttot}) and inelastic (\ref{resultinel}) cross sections.
There is an additional factor of 2 coming from extremely forward diffractive scattering which is purely elastic and have no classical analogue.

Naturally, the question of the phenomenological relevance of the obtained results arise since the
world is not $N_c=\infty$ but rather $N_c=3$.  
The total cross section predicted by the Eq. (\ref{resulttot}) is $\sigma_{\rm tot} \approx 150 \, {\rm mb}$, which
is 50\% more than the recent observation from LHC \cite{LHC} at $\sqrt{s} = 7 \, {\rm TeV}$: 
$\sigma_{\rm tot}^{\rm EXP} \approx 100 \, {\rm mb}$.
Considering the crudeness of the extreme large $N_c$ limit, the prediction for the total cross section
is in the right ballpark.
The situation for the inelastic-to-total ratio is much worse. 
The total cross section at high energies is dominated by the inelastic processes---at LHC, the ratio is 
approximately $3/4$. It is in complete disagreement with Eq. (\ref{resultratio}), since it predicts the dominant portion of
the total cross section to be elastic---the fraction of the inelastic cross section is expected to be only $1/8$.
Theoretical result does not match the observation even qualitatively.
The obvious discrepancy
between the large $N_c$ results and the observation is, however, not that surprising.
Formally, the relations (\ref{resulttot}, \ref{resultratio}, \ref{resultinel}) hold in the extreme large $N_c$ limit,
specifically for $\log(N_c) \gg 1$, which is clearly not the case for $N_c=3$. 
However, regardless of the lack of immediate phenomenological significance, the ability to calculate total cross section as well
as its inelastic component analytically is very interesting from the point of view of theoretical analysis.

\section*{Acknowledgments} 

This work was supported by the U.S.~Department of Energy
through grant DE-FG02-93ER-40762. 
V.K. also acknowledges the support of JSA/Jefferson Lab Graduate Fellowship.


\begin{thebibliography}{99}



\bibitem{tHooftNC} G. t'Hooft, Nucl. Phys. B {\bf 72} 461 (1974).
\bibitem{WittenNC} E. Witten, Nucl. Phys. B {\bf 160} 57 (1979).

\bibitem{GervaisSakita84-1} J.L. Gervais, B. Sakita, Phys. Rev. Lett. {\bf 52}, 87 (1984).

\bibitem{GervaisSakita84-2} J.L. Gervais, B. Sakita, Phys. Rev. D {\bf 30}, 1795 (1984).

\bibitem{DashenManohar93-1} R.F. Dashen, A.V. Manohar, Phys. Lett. B {\bf 315}, 425 (1993).

\bibitem{DashenManohar93-2} R.F. Dashen, A.V. Manohar, Phys. Lett. B {\bf 315}, 438 (1993).

\bibitem{DashenJenkinsManohar94} R.F. Dashen, E.E. Jenkins, A.V. Manohar, Phys. Rev. D {\bf 49}, 4713 (1994).

\bibitem{DashenJenkinsManohar95} R.F. Dashen, E.E. Jenkins, A.V. Manohar, Phys. Rev. D {\bf 51}, 3697 (1995).

\bibitem{CohenKrejcirik18} T.D. Cohen, V. Krej\v{c}i\v{r}\'ik, Phys. Rev. C {\bf 86}, 024003 (2012).

\bibitem{JenkinsLebed} E. Jenkins, R.F. Lebed, Phys. Rev. D {\bf 62} 077901 (2000).

\bibitem{Banerjee***} M.K. Banerjee, T.D. Cohen, B.A. Gelman, Phys. Rev. C {\bf 65}, 034011 (2002).

\bibitem{CohenGelman12} T.D. Cohen, B.A. Gelman, Phys. Rev. C {\bf 85} 024001 (2012).

\bibitem{CohenPRL2012} T.D. Cohen, Phys. Rev. Lett. {\bf 108} 262301 (2012).

\bibitem{CohenKrejcirik} T.D. Cohen, V. Krejcirik, Phys. Rev. C {\bf 88}, 054003 (2013).
\bibitem{KaplanSavage} D.B. Kaplan, M.J. Savage, Phys. Lett. B {\bf 365}, 244 (1996).

\bibitem{KaplanManohar} D.B. Kaplan, A.V. Manohar, Phys. Rev. C {\bf 56},  76  (1997).




\bibitem{LHC} The TOTEM Collaboration et al, EPL {\bf 96} 21002 (2011).



\end{thebibliography}
\end{document}